\documentclass{PoS}

\usepackage{verbatim}

\newcommand{\apj}{ApJ}

\newcommand{\apjs}{ApJS}

\newcommand{\aap}{A\&A}

\title{Multi-wavelength Signatures of Cosmic Rays in the Milky Way}

\ShortTitle{CRs from Multi-wavelength Observations}

\author{{\speaker{E.~Orlando}~$^a$, P.~Harrington$^a$, A.~W.~Strong$^b$} \\
\newline
\llap{$^a$}Hansen Experimental Physics Laboratory and Kavli Institute for Particle Astrophysics and Cosmology,
        Stanford University, SLAC National Accelerator Laboratory, Stanford, CA 94305, U.S.A. E-mail: \email{eorlando@stanford.edu}\\
\llap{$^b$}Max-Planck-Institut f\"ur extraterrestrische Physik, Postfach 1312, D-85741 Garching, Germany\\
        }                

\abstract{Cosmic rays (CRs) propagate in the Milky Way and interact
with the interstellar medium and magnetic fields. These interactions produce emissions that span the electromagnetic spectrum, and are an invaluable tool for
understanding the intensities and spectra of CRs in distant
regions, far beyond those probed by direct CR measurements. \\
We present updates on the study of CR properties by combining multi-frequency observations of the interstellar emission and latest CR direct measurements with propagation models.
Finally, we make predictions for e-ASTROGAM and AMEGO, proposed gamma-ray missions with a special focus at the MeV energy range.}

\FullConference{35th International Cosmic Ray Conference --- ICRC2017\\
		10--20 July, 2017\\
		Bexco, Busan, Korea}

\begin{document}

\section{Introduction}
CR direct measurements and CR indirect measurements, via observations of the interstellar emission from radio to gamma rays, are equally important in constraining CR propagation models and CR properties. 
Indeed, CRs propagate in the Galaxy, and interact with the gas in the interstellar medium (ISM) and the interstellar radiation field (ISRF) producing gamma rays via bremsstrahlung, inverse Compton scattering, and pion-decay. The same CR electrons spiraling in the magnetic field (B-field) produce synchrotron emission observed in radio and microwaves.\\
In the past years both gamma-ray and radio/microwave observations of the interstellar emission have been used to gain information on CRs together with CR direct measurements and propagation models. However, this has been done performing gamma-ray and radio analyses separately. Recent analyses and results of these works are reported in the next section. \\
Lately, there has been an ongoing effort to combine all of these observables together with the intent of obtaining more constraints on model parameters. This paper aims at giving some updates on this topic. 

\section{CR-induced interstellar emission}

In this section we summarize the latest results that regard CR-induced interstellar emissions in the context of CR propagation models. 

\subsection{The interstellar emission in gamma rays}

Recently, a detailed study of the CR-induced diffuse emission from the whole Galaxy was performed on a grid of 128 propagation models \cite{diffuse2} using the Fermi-LAT data. Even though all models provided a good agreement with data, two issues came up. First, many model-dependent structures (e.g., Fermi bubbles, Loop I, outer Galaxy) showed up as excesses over the adopted model. This was not surprising, since some of these structures were included into the models. 
Second, it was difficult to select a model that would provide the best description of the whole sky. This is likely due to the fact that some model parameters are degenerate. 
However, Fermi-LAT data hint at a large propagation halo size, additional gas in the outer Galaxy, and/or a flat CR source distribution \cite{diffuse2}. \\Further information on CRs in the Galaxy comes from observation of molecular clouds and comparison with the large-scale models based on GALPROP, e.g. \cite{Tibaldo}. New analyses of the all-sky diffuse emission are ongoing, taking advantage of model improvements and more precise observations. \\
Two accurate analyses were recently published. In one of them, the GALPROP-based models were used to analyze the spectrum and morphology of the Fermi Bubbles, the huge 10-kpc-across structures emanating from the Galactic Center \cite{Su}. The analysis revealed that both lobes have well-defined edges, and that their spectra are surprisingly hard and very similar extending up to 200 GeV \cite{bubbles}. 
The other one was devoted to the analysis of the Fermi-LAT observations of the Galactic Center \cite{GC}. This analysis yielded a weak extended residual component peaked at the Galactic Center. 
The inverse Compton component was found to be dominant and enhanced in that region. Whether this is due to CR or interstellar radiation field is still an open question. Naturally, both analyses will be repeated with updated diffuse emission models.

\subsection{The interstellar emission in radio and microwaves}
In 2011 some of us \cite{Strong2011} proposed a way to probe the local interstellar electron spectrum using high latitude total synchrotron emissivity and a collection of radio surveys and {\it WMAP}. We found that in order to reproduce the synchrotron data, the local interstellar electron spectrum should have a break at a few GeV where the value of the spectral index changes from $<$2 to 3. The break was found to be independent of propagation models. Besides, we found that the injection spectrum below a few GeV should be harder than 1.6. Plain diffusion models described the data reasonably well, while reacceleration models were less accurate at low frequencies. There, the latter were producing too many secondaries that overshot the synchrotron emission data. \\
More recently,  the total and polarized synchrotron emission were investigated for the first time in the context of physical models of CR propagation and with 3D magnetic field models  \cite{O&S2013}. Model predictions were compared with radio surveys from 22 MHz to 2.3 GHz and {\it WMAP} data at 23 GHz. After tuning the models to the Fermi all-electron measurements, we found that the all-sky total intensity and polarization maps were reasonably reproduced if an anisotropic component of the magnetic field is included; its intensity should be comparable to the regular component derived from rotation measures. This also required a flat CR distribution in the outer Galaxy and an increased size of the halo. 
The best synchrotron spectral model from \cite{O&S2013} was used for separating the low frequency components observed by {\it Planck} \cite{LFPlanck} and to provide the {\it Planck} maps officially released \cite{CompPlanck}. Indeed, at frequencies detected by {\it Planck}, the microwave sky is a superposition of different Galactic emission foregrounds that are very hard to disentangle. 
In \cite{LFPlanck} a detailed investigation of the low frequency foreground maps was performed, and some regions of the sky showed interesting structured emission.
For example, in polarization, Loop I (which was found as a filament at the same position of the edge of the northern Fermi Bubble) was found to have a harder spectral index with respect to the local spectrum and the control filament. Consequently, this suggests that the CR electrons in Loop I have a harder spectrum as well. It also suggests that the structure in polarization could be connected to the Fermi Bubbles.  However, no significant variations of the spectrum across the bubbles were found in the analysis made in \cite{bubbles}. Besides, while the bubbles are projected as emanating from the Galactic center, Loop I is not and there is no obvious counterpart in the Southern hemisphere. This suggests that Loop I and Fermi Bubbles are two different structures at different distances  \cite{LFPlanck}.\\ 
A recent work \cite{Planck_bfield} investigated different Galactic magnetic field models taken from the literature, using a representative CR distribution from \cite{O&S2013}. Intensities of the regular and random component of the Galactic magnetic field models were updated to the values that better reproduce the {\it Planck} maps.
Recently the latest AMS-02 all-electron data \cite{AMS_ele} were used to improve the model predictions for the synchrotron emission \cite{Orlando2015}. This new propagation model reproduces spectrally the synchrotron spectrum, which includes the latest {\it Planck} maps and the 408 MHz map reprocessed by \cite{408}.  

\section{Updates on the multi-frequency studies of the interstellar emission}

\begin{figure}
\center
\includegraphics[width=0.6\textwidth]{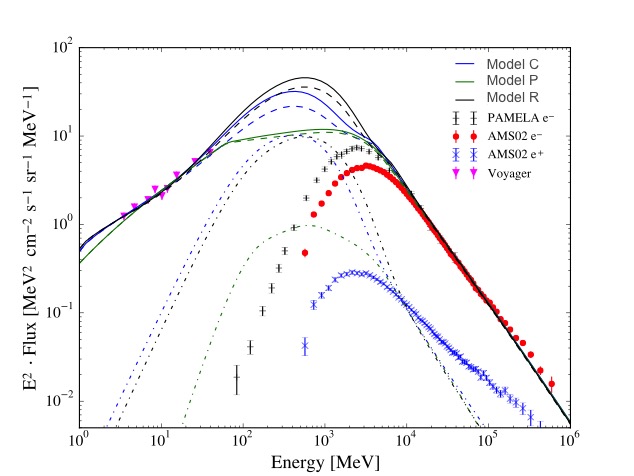}
\caption{Electron LIS of the three baseline models C (blue line), P (green line), and R (black line) for positrons (dashed-dotted line), electrons only (dashed line), and positrons plus electrons (solid line) compared with data: blue crosses: AMS-02 positrons \cite{AMS_ele}; red points: AMS-02 electrons \cite{AMS_ele}; black crosses: PAMELA electrons \cite{Pamela_ele}; magenta triangles: {\it Voyager 1} electrons and positrons \cite{Voyager}. Details on the models can be found in \cite{Orlando_inprep}}
\label{Fig1}
\end{figure}

\begin{figure}
\center
\includegraphics[width=0.6\textwidth]{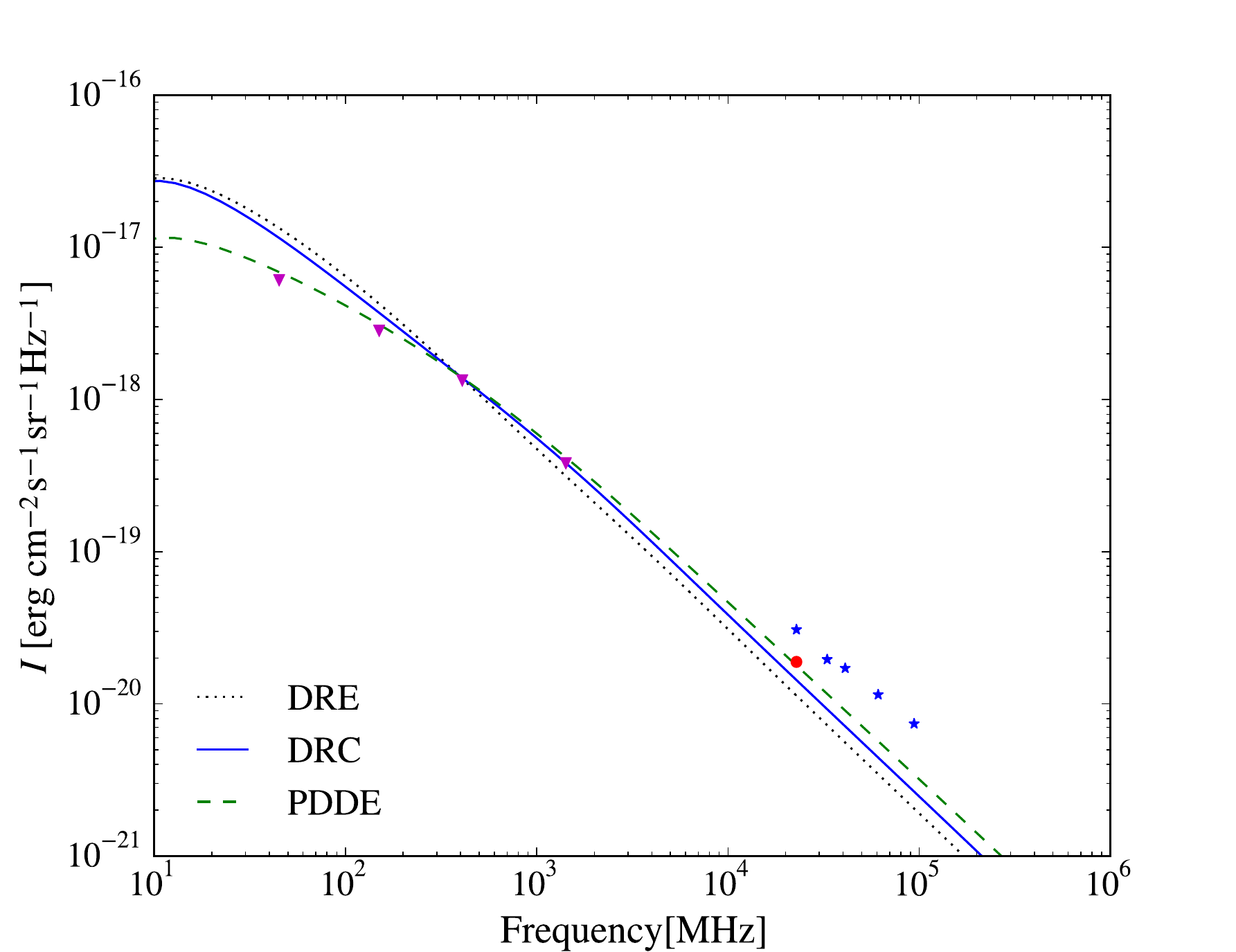}
\caption{Synchrotron spectra for the three electron models in Figure~\ref{Fig1} compared with data for intermediate latitudes (i.e. 10$^\circ$$<$b$<$20$^\circ$). Radio surveys at 45, 150, 408, and 1420\,MHz are as described in \cite{Strong2011}. {\it WMAP} \cite{Bennett} and synchrotron temperature map by {\it Planck} \cite{CompPlanck} are also shown.}
\label{Fig2}
\end{figure}

\begin{figure}
\includegraphics[width=0.32\textwidth]{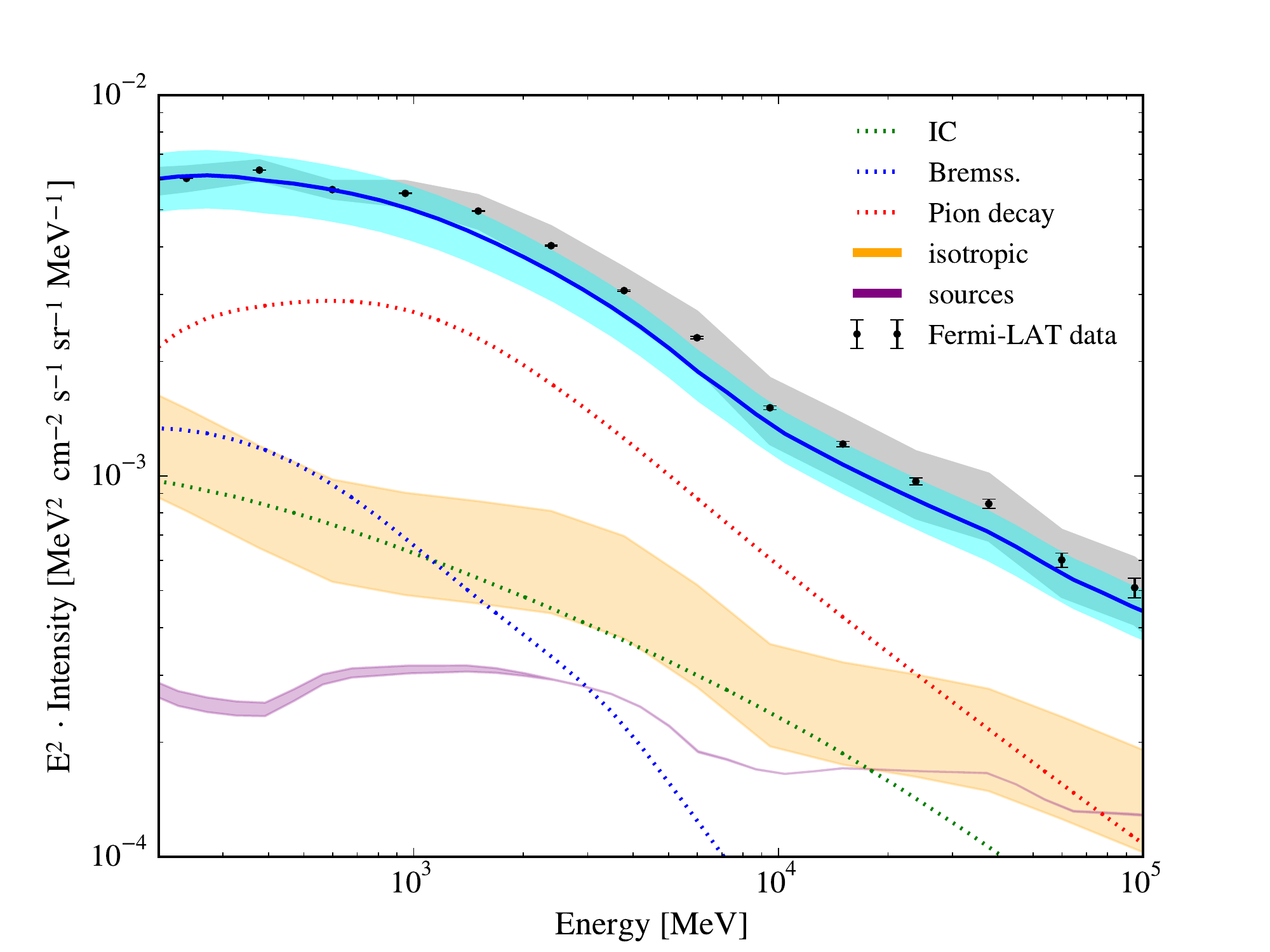}
\includegraphics[width=0.32\textwidth]{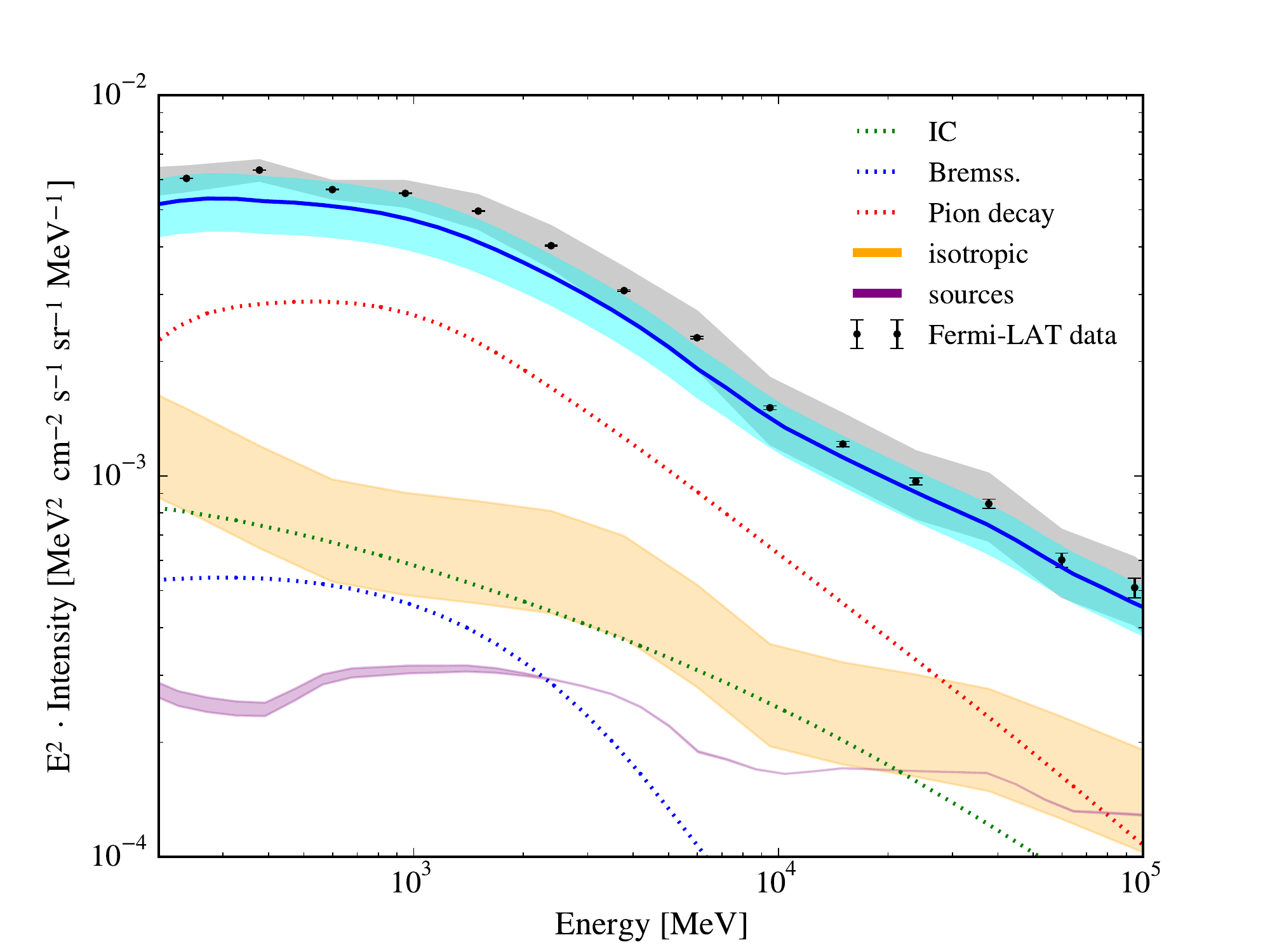}
\includegraphics[width=0.32\textwidth]{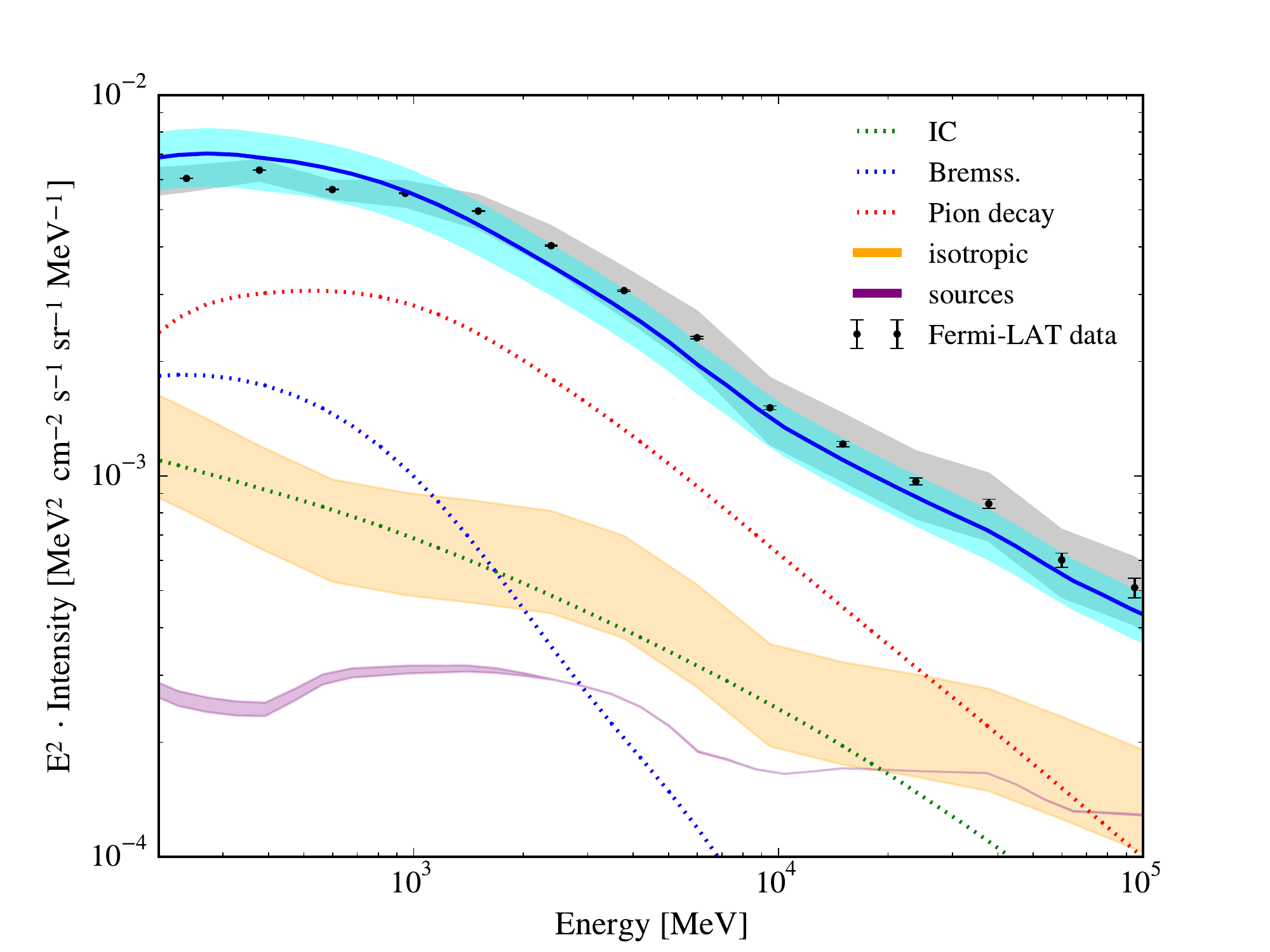}
\caption{Calculated gamma-ray components and spectrum for C, P, and R models (left to right) compared with Fermi LAT data from \cite{diffuse2} for intermediate latitudes ($10^{\circ}<|b|<20^{\circ}$, all longitudes), using the electron LIS as in Fig.\ref{Fig1} and other parameters as described in the text. Data include statistical ({\it grey area}) and systematic errors ({\it black bars}). Here components are not fitted to gamma-ray data, so uncertainties in the gas are not accounted for. Spectra for sources and isotropic are taken as in \cite{diffuse2}, for the most extreme cases reported there.  30$\%$ uncertainty is added to the isotropic spectrum, following the study in \cite{EGB} based on various foreground models. The inverse Compton component has been scaled to better reproduce gamma rays as found in \cite{diffuse2}.}
\label{Fig3}
\end{figure}

This section describes some updates to the analysis of the interstellar emission from radio to gamma rays and to the modeling of CRs with GALPROP\footnote{https://galprop.stanford.edu} \cite{Moskalenko98, Strong2007, Moskalenko2015}. 
In this paper, we focus our attention on gaining information about electrons in the energy range 10$^{2}$ - 10$^{4}$ MeV without making any assumptions about solar modulation. In this energy range solar modulation is significant, and there are no interstellar CR direct measurements. The fact that we are using radio data and {\it Voyager~1} interstellar measurements does not require any treatment of solar modulation, which illustrates the importance of using indirect observations of the interstellar emission that trace CRs.\\
First, we have updated the local interstellar spectrum (LIS) in propagation models to fit recent {\it Voyager~1} \cite{Voyager} and AMS-02 measurements \cite{AMS_ele}. Three possible CR electron spectra are shown in Figure~\ref{Fig1}. They show different electron intensities in the region 10$^{2}$ - 10$^{4}$ MeV, and are used to be tested with both synchrotron and gamma-ray observations. We report here on the spectral properties only. 
We use available interstellar emission observations in radio (radio surveys) and microwaves ({\it Planck}, {\it WMAP} data are used just for reference) at intermediate latitudes (i.e. 10$^\circ$$<$b$<$20$^\circ$) and test these electron LIS. The calculated synchrotron emission with GALPROP is shown in Figure~\ref{Fig2} for the three models. We see that the model with lower electron intensity in the range 10$^{2}$ - 10$^{4}$ MeV is the only one able to reproduce the synchrotron spectral data in the entire frequency band. The other two models where the electron intensity is significantly higher, are over-estimating the synchrotron emission at the lowest frequencies. 
At the same time, we also calculate the gamma-ray emission expected from the three models, assuming the hadronic propagation parameters as found in our previous works (i.e. \cite{Voyager} and \cite{Boschini}). Figure \ref{Fig3} shows the comparison of these predictions with Fermi LAT data for the intermediate latitudes as published in \cite{diffuse2}. 
All the models are within the Fermi LAT systematic uncertainties even without the tuning to the data that would account for uncertainties in the ISM. Hence, in a first approximation, with the data used here, all the models seems to reproduce gamma rays. 
A more detailed description of the method and results can be found in \cite{Orlando_inprep}. 

\section{Predictions for e-ASTROGAM and AMEGO}
\begin{figure}
\center
\includegraphics[width=0.7\textwidth]{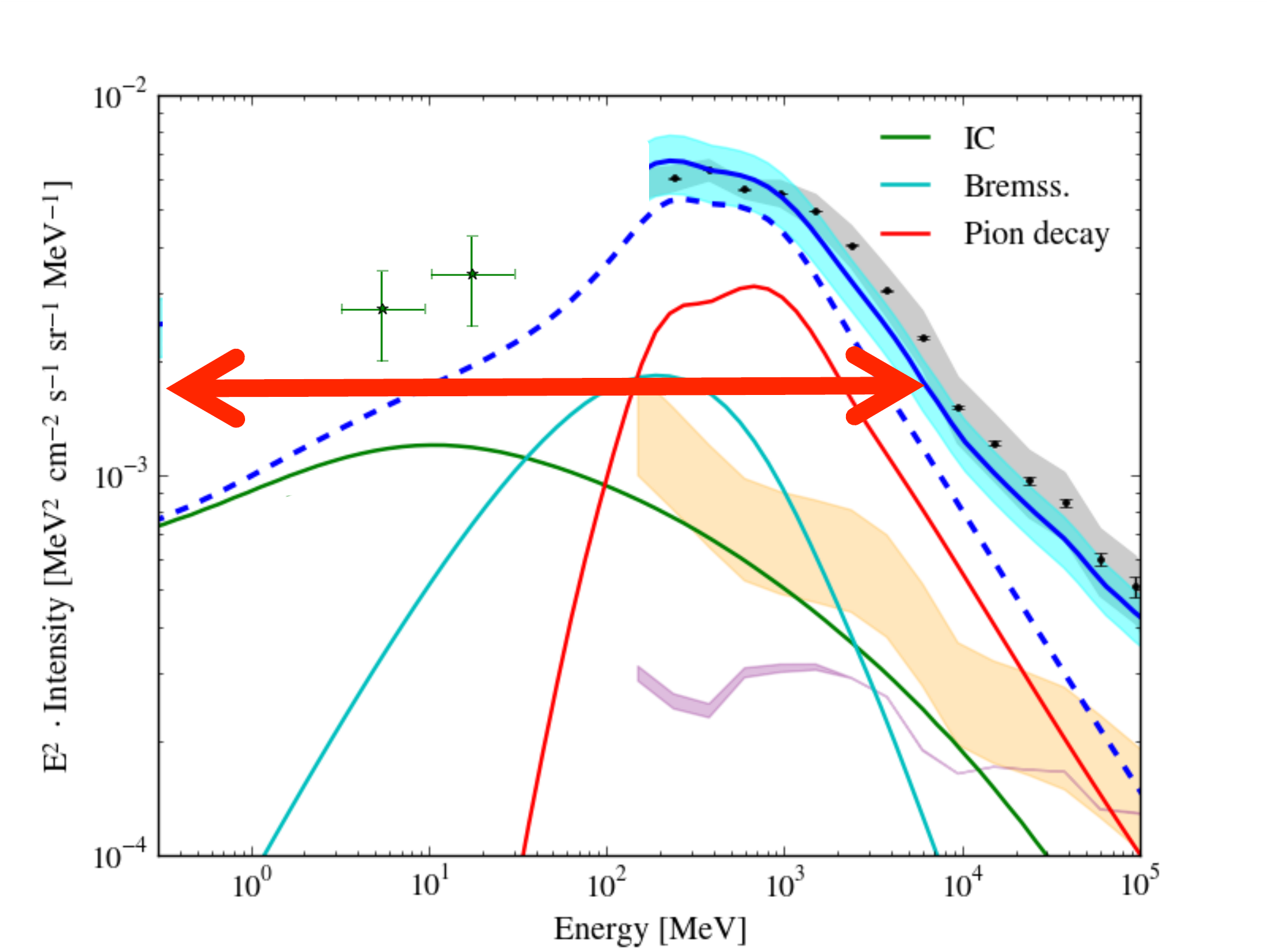}
\caption{Predictions for the energy range of e-ASTROGAM and AMEGO, possible future gamma-ray instruments for model R, as an example. The figure shows also the Fermi LAT data intermediate latitudes \cite{diffuse2} and COMPTEL data \cite{Strong1999} for comparison.}
\label{Fig4}
\end{figure}

The proposed future gamma-ray missions e-ASTROGAM \cite{eAstrogam} and AMEGO\footnote{https://asd.gsfc.nasa.gov/amego/} are designed to detect photons especially at MeV energies. In particular e-ASTROGAM is designed to detect photons from 0.3 MeV to 3 GeV, while AMEGO from 0.2 MeV to 10 GeV. In this energy range interstellar emission is a very bright component. As a consequence, different electron LIS, and hence propagation parameters, can be easily distinguished. Most importantly, this energy range covers the electron energy band that produces synchrotron emission seen in radio and microwave band. The interstellar emission below 100 MeV is supposed to be of leptonic origin, making comparisons and constraints with synchrotron emission even more important.  
Hence, we have extended the predictions of the interstellar emission down to 1 MeV for model R as an example. Figure \ref{Fig4} shows the predictions for the range of e-ASTROGAM and AMEGO.  

\section{Discussion}
In this work we show the feasibility and importance of using multi-wavelength observations, especially at radio wavelengths, together with CR measurements, to constrain the interstellar spectrum at low-energies, propagation models, and solar modulation effects. We have presented preliminary results and now discuss here the potential of this approach.

The exact derivation of the synchrotron maps as obtained by {\it Planck} and {\it WMAP} have limitations, due to the various assumptions required and degeneracies with separating multiple astrophysical components including synchrotron, free-free, thermal dust and anomalous microwave emissions (AME) (\cite{LFPlanck, CompPlanck}). Disentangling these components in the  {\it WMAP} and {\it Planck} bands requires various assumptions or a priori information such as data-driven templates using ancillary data (where each different emission mechanism is at its maximum) or by approximating their spectral properties (see \cite{Bennett, LFPlanck, CompPlanck}). 
Hence, improvements on the modeling, together with forthcoming data, could help in separating the components and may provide more strict constraints to the lepton spectrum. 

Regarding gamma rays, the sky above 100\,MeV is dominated by emission produced by CRs interacting with the gas and interstellar radiation field via pion-decay, IC, and bremsstrahlung. Disentangling the different components at the LAT energies is challenging and is usually done in a model-dependent approach. Uncertainties in the ISM is the major limitation to our modeling and hence in our knowledge of CRs, e.g. as found in \cite{diffuse2}. The situation below 100\,MeV is still unexplored. Extrapolations of present models to such low energies predict inverse Comtpon and bremsstrahlung to be the major mechanisms of CR-induced emission, which are of leptonic origin. The fact that energies $<100$\,MeV were not deeply investigated after the COMPTEL era, makes it more relevant now with the advent of the LAT Pass 8 data and its extension to lower energies \cite{Orlando, Pass8}. However disentangling the different components and characterizing the sources below 100\,MeV is even more challenging due to the relatively large point spread function of the instrument. This highlights the importance of multi-wavelength observations and the future missions at MeV energies.

\section*{Acknowledgments}
E. O. acknowledges support from NASA Grants No. NNX16AF27G.
This work makes  use of HEALPix\footnote{http://healpix.jpl.nasa.gov/} described in \cite{healpix}.

\end{document}